\documentclass{cernrep} 
\usepackage{texnames}
\usepackage[T1]{fontenc}
\usepackage[bookmarks, colorlinks=true, linktoc=page, pdftex, linkcolor=red, citecolor=red, urlcolor=red]{hyperref}
\newcommand{\be}{\begin{equation}}
\newcommand{\ee}{\end{equation}}

\begin{document}
\title{Photon--Photon Collisions with SuperChic}
 
\author{L.~A.~Harland--Lang$^a$\footnote{Speaker. Presented at PHOTON 2017 international conference on the structure and the interactions of the photon.}, V.~A.~Khoze$^{b,c}$, M.~G.~Ryskin$^c$}

\institute{$^a$ Department of Physics and Astronomy, University College London, WC1E 6BT, UK \\           
$^b$ Institute for Particle Physics Phenomenology, Durham University, DH1 3LE, UK    \\
$^c$
 Petersburg Nuclear Physics Institute, NRC Kurchatov Institute, 
Gatchina, St.~Petersburg, 188300, Russia }

\begin{abstract}
The \texttt{SuperChic} Monte Carlo generator provides a common platform for QCD--mediated, photoproduction and photon--induced Central Exclusive Production (CEP), with a fully differential treatment of soft survival effects. In these proceedings we summarise the processes generated, before discussing in more detail those due to photon--photon collisions, paying special attention to the correct treatment of the survival factor. We briefly consider the light--by--light scattering process as an example, before discussing planned extensions and refinements for the generator.
\end{abstract}


\maketitle 
 
\section{Introduction}
 
Central exclusive production (CEP) is the process
\be
pp\to p\,+\, X\, +\,p\;,
\ee
where the `+' signs indicate the presence of large rapidity gaps between the outgoing protons and the central system. That is, the protons remain intact after the collision, with just the system $X$ and nothing else (at least in the absence of pile--up) produced in the detector.  The experimental signal for this process is highly favourable, with the crucial advantage that the outgoing protons can be measured by `tagging' detectors, providing a unique insight into the properties of the central state. This has become particularly topical in light of the installation of the AFP and CT--PPS tagging detectors, which are now taking data in association with the ATLAS and CMS detectors, respectively. In addition, a novel approach based on combining the LHC Beam Loss Monitoring system with the LHC experiments may provide an alternative way to select such events~\cite{Kalliokoski:2016fjr}.

A CEP event may be produced in three ways, through a purely QCD--mediated interaction, the collision of two photons emitted from each proton, or the photoproduction process where both of these mechanisms operate. The first case  requires the development of a completely distinct theoretical framework which covers both the perturbative and non--perturbative QCD regimes, and is particularly sensitive to the nature of the produced state (see~\cite{Harland-Lang:2014lxa} for a review). In the second case, the QED initial state is particularly well understood, being simply given in terms of the known electromagnetic proton form factors, while it can be shown that the impact of non--perturbative QCD effects is small. This can therefore serve as a unique laboratory with which to observe QED mediated particle production, including of electromagnetically coupled BSM states, at the LHC. In effect, we can turn the LHC into a photon--photon collider.
 
Thus, there is a rich phenomenological CEP programme at the LHC. To fully exploit these possibilities, the \texttt{SuperChic} Monte Carlo event generator has been developed over a number of years, see~\cite{Harland-Lang:2015cta} for details.  This  simulates a broad range of exclusive processes, including QCD--mediated, those due to $\gamma\gamma$ collisions and photoproduction, and is the most complete and up-to-date generator of its kind. In these proceedings, we will discuss in detail the simulation of CEP via photon--photon collisions in \texttt{SuperChic}, demonstrating how this is achieved in the MC, describing the simulated processes, discussing one such process, namely light--by--light scattering, in more detail, before finally considering the planned extensions in the future.

\section{The \texttt{SuperChic} MC -- overview}\label{sec:over}

The \texttt{SuperChic} MC is a Fortran--based generator for CEP, providing a common platform for QCD--mediated, photoproduction and photon--induced reactions, with a fully differential treatment of soft survival effects. Arbitrary user--defined histograms may be produced, with arbitrary cuts, as well as unweighted events in Les Houches and HEPEVT formats. The code is available on the \texttt{Hepforge} website~\cite{superchic}.

 \begin{figure}
\centering\includegraphics[scale=0.9]{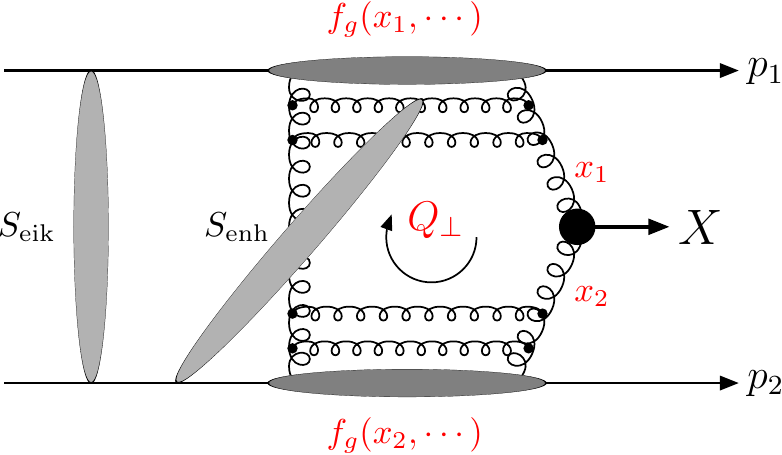}\qquad\qquad
\centering\includegraphics[scale=1.0]{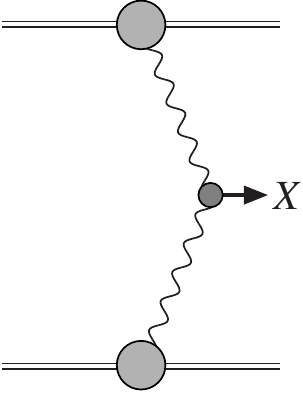}
\caption{The mechanisms for QCD (Left) and (Right) photon--initiated CEP.}
\label{fig:CEP}
\end{figure}

\subsection{QCD--initiated processes}

A representative diagram of the QCD--initiated mechanism for CEP is shown in Fig.~\ref{fig:CEP} (Left). The pQCD--based Durham model is used to calculate the basic cross section, while the model of~\cite{Khoze:2013dha} is used to include survival effects, that is the probability that additional soft particle production will not spoil the exclusivity of the event (the same model is applied to the photon--initiated and photoproduction processes listed in Sections~\ref{sec:photgen} and~\ref{sec:photoprod}). The processes currently generated are the production of:

\begin{itemize}

\item Standard Model Higgs boson via the $b\overline{b}$ decay.
\item 2-- and 3--jet events.
\item Light meson pairs ($\pi\pi$, $\eta(')\eta(')$, $KK$, $\phi\phi$, $\rho\rho$).
\item Quarkonium pairs ($J/\psi$, $\psi(2S)$)
\item $\chi_{c,b(J)}$ quarkonia, via 2-- and 3--body decays, and $\eta_{c,b}$.
\item Photon pairs, $\gamma\gamma$.

\end{itemize}

\subsection{Photon--initiated processes}\label{sec:photgen}

The basic CEP process is shown in Fig.~\ref{fig:CEP} (Right) and is discussed in more detail in the following sections. Here we simply list the generated processes, which are available for proton and lepton beams:

\begin{itemize}

\item Standard Model Higgs boson via the $b\overline{b}$ decay.
\item $W^+W^-$ via leptonic decays, including full spin correlations.
\item Lepton pairs, $l^+l^-$.
\item Light--by--light scattering, $\gamma\gamma$.
\item Monopolium and Monopole pairs\footnote{Available on request but not in official release at the time of these proceedings.}.

\end{itemize}

\subsection{Photoproduction}\label{sec:photoprod}

These are simulated in the MC according to a fit to the available HERA data. The generated final--states are:

\begin{itemize}

\item $\rho(\to \pi^+\pi^-)$.
\item $\phi(\to K^+K^-)$.
\item $J/\psi (\to \mu^+\mu^-)$.
\item $\Upsilon (\to \mu^+\mu^-)$.
\item $\psi(2S)(\to \mu^+\mu^-, J/\psi \pi^+\pi^-$).

\end{itemize}

\section{Modelling $\gamma\gamma$ collisions}

Exclusive photon--exchange processes in $pp$ collisions are described in terms of the equivalent photon approximation (EPA)~\cite{Budnev:1974de}. The quasi--real photons are emitted by the incoming proton $i=1,2$ with a flux given by
\begin{equation}\label{WWflux}
n(x_i)=\frac{\alpha}{\pi x_i}\int\frac{{\rm d}^2p_{i_\perp} }{p_{i_\perp}^2+x_i^2 m_p^2}\left(\frac{p_{i_\perp}^2}{p_{i_\perp}^2+x_i^2 m_p^2}(1-x_i)F_E(Q_i^2)+\frac{x_i^2}{2}F_M(Q_i^2)\right)\;,
\end{equation}
where $x_i$ is photon momentum fraction, and $Q^2$ is the photon virtuality, given by
\begin{equation}\label{qi}
Q^2_i=\frac{p_{i_\perp}^2+x_i^2 m_p^2}{1-x_i}\;.
\end{equation}
The functions $F_E$ and $F_M$ are given in terms of the proton electric and magnetic form factors, and are known in the phenomenologically relevant region to sub--percent level precision; we take the `double--dipole' parameterisation as measured by the A1 collaboration~\cite{Bernauer:2013tpr}. The CEP cross section is then given in terms of the photon flux by
\be\label{eq:bare}
\frac{{\rm d} \sigma^{pp\to pXp}}{{\rm d}M_X^2\,{\rm d}y_X}\sim \frac{1}{s}n(x_1)n(x_2)\cdot \hat{\sigma}(\gamma\gamma\to X)\;,
\ee
where $M_X$, $Y_X$ are the mass and rapidity of the produced object $X$. According to a naive application of the EPA this would in fact be an exact equality. However, this omits the fact that we are asking for an exclusive final--state, that is the production of $X$ accompanied by no additional particles. In addition to the photon--initiated interaction above, the protons may interact independently, producing soft particles and spoiling the exclusivity of the final state. In other words, for an exclusive process we must include the probability of no multi--parton interactions, known as the `survival factor'. 

The inclusion of the survival factor in a MC environment requires some care, as this is not a simple multiplicative factor, but rather it depends on the final--state kinematics. To see why this is the case we note that the probability for additional particle production must depend on physical grounds on the impact parameter of the colliding protons; most simply, if the protons collide at larger impact parameter they will be less likely to interact independently and produce additional particles. More concretely, the survival factor obeys
\be \label{eq:s2}
\frac{{\rm d} S^2}{{\rm d}^2 {\mathbf b}_{1t}{\rm d}^2 {\mathbf b}_{2t}} \sim |T\left(s, {\mathbf b}_{1t}, {\mathbf b}_{2t}\right)|^2\,\exp (-\Omega(s, {\mathbf b}_{t}))\;,
\ee
where ${\mathbf b}_{it}$ are the transverse positions of the colliding protons, and ${\mathbf b}_{t}={\mathbf b}_{1t}-{\mathbf b}_{2t}$ is the impact parameter of the collision. $T$ is the  amplitude corresponding to the cross section (\ref{eq:bare}) excluding survival effects, and $\Omega$ is the proton opacity, which relates to the non--perturbative structure of the proton, and can be extracted from such hadronic observables as the elastic and total cross sections. Thus $\exp (-\Omega)$ corresponds to the Poissonian probability for no additional particle production.

In the MC we do not work explicitly in impact parameter space, but rather (\ref{eq:s2}) translates into a dependence on the transverse momenta ${\mathbf p}_{i\perp}$ of the outgoing protons, which are the Fourier conjugates of the ${\mathbf b}_{it}$ variables. By considering the photon flux (\ref{WWflux}) this allows us to make some immediate conclusions about the impact of the survival factor. In particular, due to the form factors $F_E$, $F_M$, which are steeply falling with photon virtuality, the average $Q^2_i\sim p_{i\perp}^2\sim 0.05$ ${\rm GeV}^2$ is very low in photon--initiated CEP. This corresponds to large impact parameters $\gtrsim 1$ fm and therefore $S^2 \sim 1$. In other words, the impact of non--perturbative QCD effects is low, and to good approximation we are dealing with a purely QED initial state. This supports the use of such processes as tools to search for electromagnetically charged BSM states. 

\begin{figure}
\centering\includegraphics[scale=0.45]{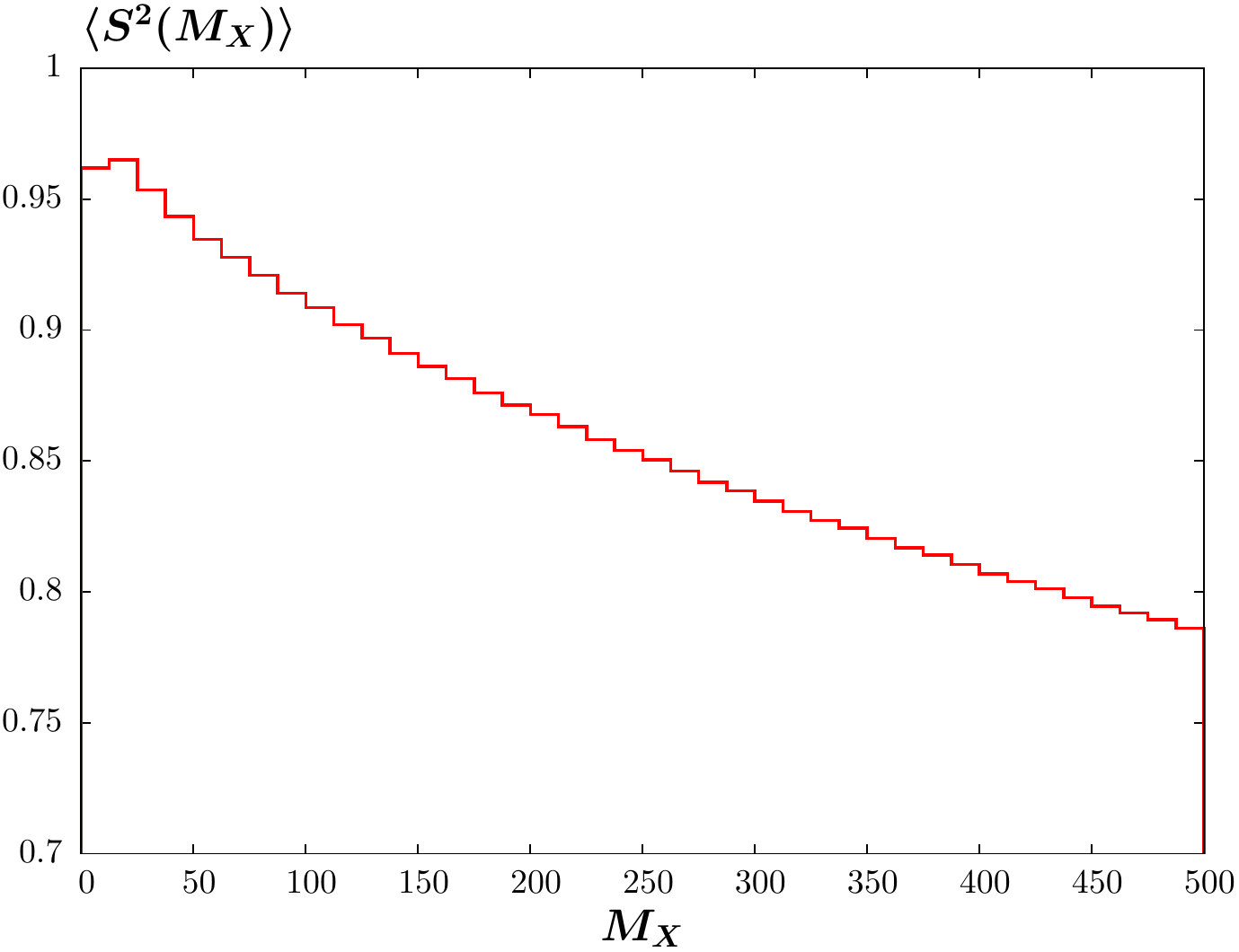}\qquad
\includegraphics[scale=0.6]{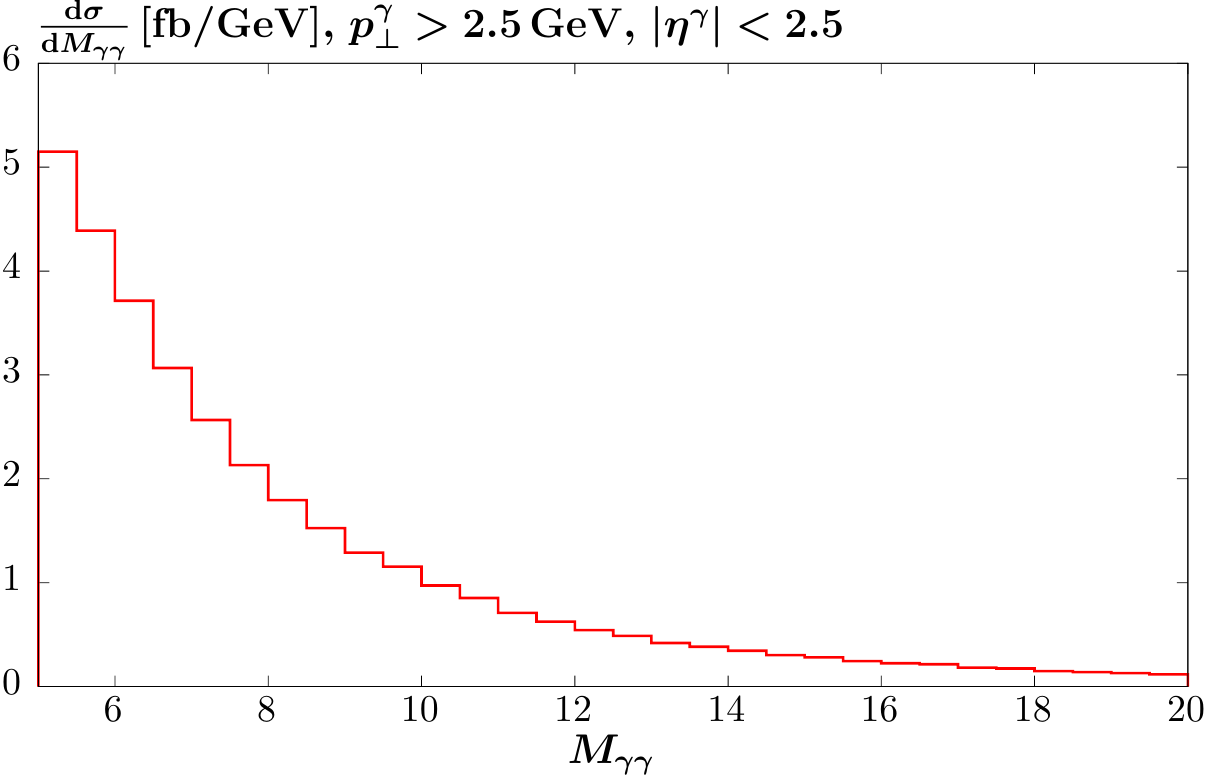}
\caption{(Left) Average survival factor for lepton pair production at the 14 TeV LHC, taken from~\cite{Harland-Lang:2015cta}. The leptons are required to have $p_\perp>2.5$ GeV and $|\eta|<2.5$. (Right) $\gamma\gamma$ invariant mass distribution for light--by--light scattering in $pp$ collisions at the 13 TeV LHC. The photons are required to have $p_\perp>2.5$ GeV and $|\eta|<2.5$. Both distributions calculated using \texttt{SuperChic}.}
\label{fig:s2m}
\end{figure}

A further implication of this derives from (\ref{qi}), from which we can see that the average photon virtuality will increase with increasing momentum fraction $x$, and therefore we will expect the survival factor to decrease as the invariant  mass and/or rapidity of the produced object increases. This trend is clear in Fig.~\ref{fig:s2m} (Left), which shows the dependence of the average survival factor on the invariant mass $M_X$ of produced system for the case of lepton pair production at the 14 TeV LHC. In \texttt{SuperChic}, a complete differential treatment of the survival factor is provided, so that all such kinematic effects are automatically accounted for. In fact, while (\ref{eq:bare}) and (\ref{eq:s2}) are written at the cross section level, a proper treatment of survival effects requires that we work at the amplitude level. In this way, the survival factor is sensitive to the helicity structure of the underlying $\gamma\gamma \to X$ subprocess, see~\cite{Harland-Lang:2015cta} for further details. 

It is worth emphasising that the impact parameter dependence of both the opacity $\Omega$ and the $\gamma\gamma \to X$ amplitude $T$ in (\ref{eq:s2}) must be accounted for, and if this is omitted it will give misleading results. This is the case in for example~\cite{Dyndal:2014yea}, which has been compared to the recent ATLAS measurement~\cite{Aaboud:2017oiq} of exclusive muon pair production. The principle cause for the difference between these results and the \texttt{SuperChic} prediction is not the choice of model for the opacity $\Omega$ (which may have some genuine model variation) but rather the fact that the impact parameter dependence of the $\gamma\gamma\to \mu^+\mu^-$ amplitude is omitted in~\cite{Dyndal:2014yea}. This has been checked explicitly in~\cite{Harland-Lang:2015cta}.

\section{Example process: light--by--light scattering}\label{sec:lbyl}

The full list of generated photon--initiated CEP processes is given in Section~\ref{sec:photgen}. As an example, we will consider the case of light--by--light scattering, $\gamma\gamma \to \gamma\gamma$, where in the SM the continuum process proceeds via an intermediate lepton, quark and $W$ boson box, see~\cite{dEnterria:2013zqi,Klusek-Gawenda:2016euz} for a detailed study. Until recently, this process had not been observed directly, and it is also sensitive to BSM effects, see e.g.~\cite{Ellis:2017edi,Knapen:2016moh}. In addition, it is particularly topical in light of the first direct evidence for this process by the ATLAS collaboration~\cite{Aaboud:2017bwk}, in Pb--Pb collisions. The invariant mass distribution for the 13 TeV in $pp$ collisions is shown in Fig.~\ref{fig:s2m} (Right). 

While in the official \texttt{SuperChic} release only lepton and proton beams are available, a version with the heavy ion flux implemented using code provided by the authors of~\cite{dEnterria:2013zqi} is available on request. Work is currently ongoing to include heavy ion beams in the MC, including an exact treatment of the initial--state kinematics and a proper evaluation of survival effects.

In addition to the light--by--light signal, it is in general possible for the exclusive $\gamma\gamma$ final--state to be produced via the QCD interaction $gg\to \gamma\gamma$ as in Fig.~\ref{fig:CEP} (Left). In the ATLAS analysis, to estimate the $Pb$--$Pb$ cross section the \texttt{SuperChic} prediction for $pp$ collisions is corrected by a factor of $A^2 R_g^2$, taken from~\cite{dEnterria:2013zqi}, where $A=208$ is the lead mass number and $R_g \approx 0.7$ accounts for nuclear shadowing effects. In other words, up to the shadowing correction the predicted cross section in $pp$ collisions is simply scaled by the number of participating nucleons in the collision. However, this argument is not justified. In particular, as the range of QCD $R_{\rm QCD}\ll R_A$ only those nucleons which are situated on the ion periphery may interact while leaving the ions intact. A detailed calculation is therefore required, with the survival factor evaluated by correctly accounting for the geometry of the heavy ion collision. In this way, we find that the CEP cross section in heavy ion collisions will instead scale like $\sim A^{1/3}$~\cite{Harland-Lang:fut}. We will therefore expect the QCD--initiated contribution to be lower than a simple $A^2$ scaling would suggest, although a precise numerical prediction is required to confirm the level of suppression\footnote{In fact in~\cite{Aaboud:2017bwk} the normalization of the \texttt{SuperChic} result is determined by the data, giving a value of $f_g=0.5\pm 0.3$ relative to the prediction. This however is driven by a limited number of events in the tail of the photon acoplanarity distribution which may also be due to dissociation of the ions; as the ZDCs were not used in the ATLAS analysis, this cannot be excluded.}.

\section{Conclusion and outlook}

The \texttt{SuperChic} Monte Carlo generator provides a common platform for QCD--mediated, photoproduction and photon--induced reactions, with a fully differential treatment of soft survival effects. The latest version is available  on the \texttt{Hepforge} website~\cite{superchic}. In these proceedings we have concentrated on the case of photon--photon collisions, but the full list of the generated processes has been given in Section~\ref{sec:over}.

A number of extensions to the MC are planned for the future. There are two general possibilities to pursue, either including new beam types or adding new processes, and extensions in both directions are foreseen. In the former case, as discussed in Section~\ref{sec:lbyl}, work is ongoing to include a complete treatment of heavy ion beams for photon--induced processes, as well as a precise evaluation of the QCD--initiated cross section. In the latter case, a range of additions are anticipated for the next MC version, including axion--like particles, monopolium and monopole pairs and the inclusion of $W$ boson loops in the case of light--by--light scattering (which can be important at higher masses, and is so far omitted). By continuing to develop and extend this tool, we hope to exploit as fully as possible the exciting potential to use the LHC as photon--photon collider at unprecedented energies.

\section*{Acknowledgements}

MGR thanks the IPPP at Durham University for hospitality and VAK thanks the Leverhulme Trust for an Emeritus Fellowship. The research of MGR was supported by the RSCF grant 14-22-00281. LHL thanks the Science and Technology Facilities Council (STFC) for support via the grant award ST/L000377/1.


\begin{thebibliography}{99}

  
\bibitem{Kalliokoski:2016fjr}
  M.~Kalliokoski, J.~W.~Lämsä, M.~Mieskolainen and R.~Orava,
  arXiv:1604.05778 [hep-ex].


\bibitem{Harland-Lang:2014lxa}
  L.~A.~Harland-Lang, V.~A.~Khoze, M.~G.~Ryskin and W.~J.~Stirling,
  Int.\ J.\ Mod.\ Phys.\ A {\bf 29} (2014) 1430031
  doi:10.1142/S0217751X14300312
  [arXiv:1405.0018 [hep-ph]].
  
\bibitem{Harland-Lang:2015cta}
  L.~A.~Harland-Lang, V.~A.~Khoze and M.~G.~Ryskin,
  Eur.\ Phys.\ J.\ C {\bf 76} (2016) no.1,  9
  doi:10.1140/epjc/s10052-015-3832-8
  [arXiv:1508.02718 [hep-ph]].

\bibitem{superchic}
  {\tt http://projects.hepforge.org/superchic/}
	
\bibitem{Khoze:2013dha}
  V.~A.~Khoze, A.~D.~Martin and M.~G.~Ryskin,
  Eur.\ Phys.\ J.\ C {\bf 73} (2013) 2503
  doi:10.1140/epjc/s10052-013-2503-x
  [arXiv:1306.2149 [hep-ph]].
	
\bibitem{Budnev:1974de}
  V.~M.~Budnev, I.~F.~Ginzburg, G.~V.~Meledin and V.~G.~Serbo,
  Phys.\ Rept.\  {\bf 15} (1975) 181.
  doi:10.1016/0370-1573(75)90009-5
	
\bibitem{Bernauer:2013tpr}
  J.~C.~Bernauer {\it et al.} [A1 Collaboration],
  Phys.\ Rev.\ C {\bf 90} (2014) no.1,  015206
  doi:10.1103/PhysRevC.90.015206
  [arXiv:1307.6227 [nucl-ex]].
	

  
 
\bibitem{Dyndal:2014yea}
  M.~Dyndal and L.~Schoeffel,
  Phys.\ Lett.\ B {\bf 741} (2015) 66
  doi:10.1016/j.physletb.2014.12.019
  [arXiv:1410.2983 [hep-ph]].

\bibitem{Aaboud:2017oiq}
  M.~Aaboud {\it et al.} [ATLAS Collaboration],
  arXiv:1708.04053 [hep-ex].

\bibitem{dEnterria:2013zqi}
  D.~d'Enterria and G.~G.~da Silveira,
  Phys.\ Rev.\ Lett.\  {\bf 111} (2013) 080405
   Erratum: [Phys.\ Rev.\ Lett.\  {\bf 116} (2016) no.12,  129901]
  doi:10.1103/PhysRevLett.111.080405, 10.1103/PhysRevLett.116.129901
  [arXiv:1305.7142 [hep-ph]].
  
  \bibitem{Klusek-Gawenda:2016euz} 
  M.~Klusek-Gawenda, P.~Lebiedowicz and A.~Szczurek,
  Phys.\ Rev.\ C {\bf 93}, no. 4, 044907 (2016)
  doi:10.1103/PhysRevC.93.044907
  [arXiv:1601.07001 [nucl-th]].


\bibitem{Ellis:2017edi}
  J.~Ellis, N.~E.~Mavromatos and T.~You,
  Phys.\ Rev.\ Lett.\  {\bf 118} (2017) no.26,  261802
  doi:10.1103/PhysRevLett.118.261802
  [arXiv:1703.08450 [hep-ph]].

\bibitem{Knapen:2016moh}
  S.~Knapen, T.~Lin, H.~K.~Lou and T.~Melia,
  Phys.\ Rev.\ Lett.\  {\bf 118} (2017) no.17,  171801
  doi:10.1103/PhysRevLett.118.171801
  [arXiv:1607.06083 [hep-ph]].
  
\bibitem{Aaboud:2017bwk}
  M.~Aaboud {\it et al.} [ATLAS Collaboration],
  arXiv:1702.01625 [hep-ex].
  
\bibitem{Harland-Lang:fut}
  L.~A.~Harland-Lang, V.~A.~Khoze and M.~G.~Ryskin,
  Future Publication.

\end{thebibliography}
\end{document}